\documentclass[aps,prd,eqsecnum,floats,floatfix,twocolumn,showpacs]{revtex4}

\usepackage{graphicx}
\usepackage{bm}

\newcommand{\anp}{g_{np}}
\newcommand{\anS}{g_{n\Sigma}}
\newcommand{\apL}{g_{p\Lambda}}
 
\newcommand{\manp}{\left(1-\anp^2\right)}
\newcommand{\panp}{\left(1+\anp^2\right)}
\newcommand{\mapL}{\left(1-\apL^2\right)}
\newcommand{\papL}{\left(1+\apL^2\right)}
\newcommand{\manS}{\left(1-\anS^2\right)}
\newcommand{\panS}{\left(1+\anS^2\right)}
 
\newcommand{\en}{\epsilon_n}
\newcommand{\ep}{\epsilon_p}
\newcommand{\eL}{\epsilon_\Lambda}
\newcommand{\pL}{p_\Lambda}
\newcommand{\eS}{\epsilon_\Sigma}
\newcommand{\pS}{p_\Sigma}

\begin{document}

\title{Effect of hyperon bulk viscosity on neutron-star $r$-modes}

\author{Lee Lindblom$^1$ and Benjamin J. Owen$^{2,3}$}

\affiliation{${}^1$ Theoretical Astrophysics 130-33, California Institute of
Technology, Pasadena, CA 91125}
\affiliation{${}^2$ Department of Physics, University of Wisconsin--Milwaukee,
P.O. Box 413, Milwaukee, WI 53201}
\affiliation{${}^3$ Albert Einstein Institut (Max Planck Institut f\"ur
Gravitationsphysik), Am M\"uhlenberg 1, 14476 Golm, Germany}

\begin{abstract}
Neutron stars are expected to contain a significant number of hyperons
in addition to protons and neutrons in the highest density portions of
their cores.  Following the work of Jones, we calculate the
coefficient of bulk viscosity due to nonleptonic weak interactions
involving hyperons in neutron-star cores, including new relativistic
and superfluid effects.  We evaluate the influence of this new bulk
viscosity on the gravitational radiation driven instability in the
$r$-modes.  We find that the instability is completely suppressed in
stars with cores cooler than a few times $10^9$~K, but that stars
rotating more rapidly than $10-30\%$ of maximum are unstable for
temperatures around $10^{10}$~K.  Since neutron-star cores are
expected to cool to a few times $10^9$~K within seconds (much shorter
than the $r$-mode instability growth time) due to direct Urca
processes, we conclude that the gravitational radiation instability
will be suppressed in young neutron stars before it can significantly
change the angular momentum of the star.
\end{abstract}

\pacs{97.60.Jd, 04.40.Dg, 26.60.+c, 04.30.Db}

\maketitle

\section{Introduction}

The $r$-modes (fluid oscillations whose dynamics is dominated by
rotation) of neutron stars have received considerable attention in the
past few years because they appear to be subject to the
Chandrasekhar-Friedman-Schutz gravitational radiation instability in
realistic astrophysical conditions (see Ref.~\cite{lindblom} for a
recent review). If the $r$-modes are unstable, i.e.\ if the damping
timescales due to viscous processes in neutron-star matter are longer
than the gravitational-radiation driving timescale, a rapidly rotating
neutron star could emit a significant fraction of its rotational
energy and angular momentum as gravitational waves.  With appropriate
data analysis strategies, these waves could be detectable by interferometers
comparable to enhanced
LIGO. The $r$-mode instability might also explain the
relatively long spin periods observed in young pulsars and of older,
accreting pulsars in low-mass x-ray binaries.
 
Recently Jones~\cite{jones01,jones2} has pointed out that long-neglected
processes involving hyperons (massive cousins of the nucleons) can
lead to an extremely high coefficient of bulk viscosity in the core of
a neutron star.  Using simple scaling arguments he suggests that the
viscous damping timescales associated with these processes may be
short enough to suppress the $r$-mode instability altogether in
realistic astrophysical circumstances.  The purpose of this paper is
to investigate this possibility more thoroughly.  The hyperons only
exist in the central core of a neutron star where the density is
sufficiently high.  The relevant effects of the $r$-modes however
vanish as $r^6$ (where $r$ is the distance from the center of the
star).  Thus the overall effect of hyperon induced dissipation on the
$r$-modes depends very sensitively on the size and structure of the
core of a neutron star.  Jones' initial estimates did not take
properly into account either the structure of the $r$-mode or the
detailed properties of the nuclear matter in the core of a neutron
star.  We improve on Jones' analysis in several ways: First we
evaluate fully relativistic cross sections to determine the reaction
rates of the relevant hyperon interactions.  We find that these cross
sections reduce to the results of Jones~\cite{jones01,jones71} in the
low-momentum limit, but can be about an order of magnitude larger in some
regimes of neutron-star matter.  Second we derive new expressions
for the bulk viscosity coefficient that are appropriate even for a
relativistic fluid such as neutron star matter.  Third we construct
detailed neutron star models based on an equation of state that
includes hyperons and the appropriate interactions among all of the
particle species present.  Due to superfluid effects the temperature
and density dependence of hyperon bulk viscosity turns out to be quite
complicated: superfluidity increases the viscosity in some cases while
reducing it in others.  And fourth, we use a more accurate model of the
structure of the $r$-mode eigenfunction in the cores of these stars to
evaluate the effects of hyperon dissipation.

Our analysis shows that hyperon bulk viscosity completely suppresses the
gravitational radiation instability in the $r$-modes of rotating neutron
stars for temperatures below a few times $10^9$~K.  We find that the
gravitational radiation instability acts most strongly at temperatures
around $10^{10}$~K where stars rotating more than $10-30\%$ of the maximum
rotation rate (depending on the details of the microphysics) are driven
unstable.  Our coefficient of bulk viscosity is actually several hundred
times that of Jones~\cite{jones01,jones2}, who suggested that the
instability was completely suppressed. However, our use of the proper
$r$-mode eigenfunction reduces the dissipation by several orders of
magnitude and we find that there is a window of instability. How long it
lasts is another matter.  If the core of the neutron star cools via the
standard modified Urca process, its temperature remains above a few times
$10^9$~K for about a day~\cite{shapiro-teuk}.  This is enough time for an
unstable $r$-mode to grow and radiate away a substantial fraction of the
star's rotational kinetic energy and angular momentum into gravitational
waves~\cite{ltv}.  However if the core of the star cools too quickly the
instability might not have enough time to grow before being suppressed
by the hyperon bulk viscosity.  The time needed for a neutron-star
core to cool to a few times $10^9$~K is reduced to about a second when
direct Urca processes are able to act~\cite{pt,page-etal}.  Modern equations
of state have large enough proton densities in the core that direct Urca
cooling is now expected to act until the neutrons and protons condense into
a superfluid state, i.e. above about $10^9$~K.  The growth time for the
gravitational radiation instability in the most rapidly rotating neutron
stars is about 40~s~\cite{lom98}.  Thus we conclude that the core of a
neutron star will probably cool too quickly for the $r$-mode instability to
grow significantly before being suppressed by the hyperon bulk
viscosity.

The organization of the rest of this paper is as follows. In
Sec.~\ref{s:eos} we provide details of the equation of state which we
use, including numerical aspects of the evaluation of various
thermodynamic variables and derivatives, and the model of hyperon
superfluidity that we employ.  In Sec.~\ref{s:bulk} we present a new
derivation of the coefficient of bulk viscosity for relativistic
neutron-star matter (including several interacting fluids) in terms of
the microscopic reaction rates and thermodynamic derivatives.  In
Sec.~\ref{s:rates}, we compute the relevant cross sections in order to
evaluate the reaction rates for hyperons in a dense medium. In
Sec.~\ref{s:times} we combine the thermodynamic expressions of
Sec.~\ref{s:bulk} with the microscopic reaction rates of
Sec.~\ref{s:rates} to obtain expressions for hyperon bulk viscosity as
a function of density and temperature in neutron-star matter.  In
Sec.~\ref{s:rmodes} we evaluate $r$-mode damping timescales for
neutron stars containing ordinary fluid and superfluid hyperons.
Finally, in Sec.~\ref{s:end} we discuss the implications of
our results for the $r$-mode instability in real neutron stars, and
also attempt to estimate how robust these conclusions are.

\section{Equation of state}
\label{s:eos}
\subsection{Thermodynamic Equilibrium}
Neutron-star matter is a Fermi liquid which at low densities is
composed primarily of neutrons $n$, protons $p$ and electrons $e$.
Charge neutrality $n_p=n_e$ (where $n_i$ is the number density of the
$i^{\rm\, th}$ species) and $\beta$-equilibrium $\mu_n=\mu_p+\mu_e$
(where $\mu_i$ is the chemical potential of the $i^{\rm\, th}$
species) determine the relative concentrations of these particles at
each density.  As the total baryon density increases however, it
becomes energetically favorable for the equilibrium state to include
additional particle species: first muons $\mu$, and then a sequence of
hyperons $\Sigma^-$, $\Lambda$, ...  These additional particles appear
as the density exceeds the threshold for the creation of each new
species.  The relative concentrations of the various species are
determined at each density by imposing charge neutrality and
$\beta$-equilibrium.  At the highest densities of interest to us these
equilibrium constraints are
\begin{eqnarray}
n_p&=&n_e+n_\mu+n_{\Sigma^-},\\
\mu_p&=&\mu_n-\mu_e,\\
\mu_\mu&=&\mu_e,\\
\mu_{\Sigma^-}&=&\mu_n+\mu_e,\\
\mu_\Lambda&=&\mu_n.
\end{eqnarray}

\begin{figure}
\includegraphics[width=3.0in]{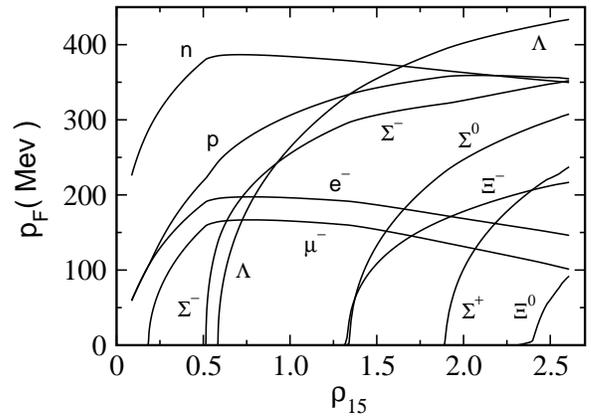}
\caption{Fermi momenta (in MeV) of the baryons and leptons as functions of
total energy density (in $10^{15}$g/cm$^3$) for Glendinning's equation of
state.\label{f:p_fermi}}
\end{figure}

In order to solve these constraints and determine the equilibrium
state of neutron-star matter, we need explicit expressions for the
various chemical potentials $\mu_i$ as functions of the particle
number densities $n_j$.  These functions have encoded within them the
details of the interactions between the various particles in a dense
Fermi-liquid environment.  In this paper we have adopted the
expressions for these chemical potentials as given by Glendenning's
relativistic effective mean-field theory~\cite{glendpaper,glendbook}.
Figure~\ref{f:p_fermi} illustrates the Fermi momenta of the various
particle species as a function of the total energy density of the
matter that we obtained with Glendenning's (case 2~\cite{glendpaper})
expressions for the chemical potentials.  Glendenning also gives
expressions for the total energy density $\rho$ and total pressure $p$
as functions of the particle densities $n_i$.  These quantities are
illustrated and tabulated by Glendenning~\cite{glendpaper,glendbook},
and we will not reproduce them here.  Our numerical code reproduces
Glendenning's numbers quite accurately.

We are also interested here in some less familiar thermodynamic
quantities that are relevant for calculating the bulk viscosity in
neutron-star matter.  These quantities are easily determined once the
full description of the equilibrium state is known.  In particular the
partial derivatives of the chemical potentials with respect to the
various particle number densities, $\alpha_{ij}\equiv
\partial\mu_i/\partial n_j$, are needed in the expression for the
relaxation time associated with bulk visocity as defined in
Eqs.~(\ref{tau}), (\ref{delta1}), and (\ref{delta2}) below.  These
$\alpha_{ij}$ are easily determined numerically (or even analytically
in some cases) once the full equilibrium state is known.  Further the
thermodynamic function,
\begin{equation}
\gamma_\infty-\gamma_0\equiv -{n_B^2\over p}
{\partial p\over \partial n_n}{d\tilde
x_n\over dn_B},\label{prefactor}
\end{equation}
appears as a prefactor in the expression for the bulk viscosity,
Eq.~(\ref{zeta}), that we derive below.  Here ${\partial p/\partial
n_n}$ is just the partial derivative of the pressure with respect to
the number density of neutrons (keeping the other number densities
fixed), and $d\tilde x_n/d n_B$ is the derivative of the fractional
density of neutrons in the equilibrium state, $\tilde x_n=n_n/n_B$,
with respect to the total baryon density $n_B$.  The left side of
Eq.~(\ref{prefactor}) has been re-expressed in terms of
$\gamma_\infty$ the ``fast'' and $\gamma_0$ the ``slow'' adiabatic
indices defined in Eqs.~(\ref{gammai}) and (\ref{gamma0}) below.
Figure~\ref{f:thermo} illustrates this function for the Glendenning
equation of state.  For a non-relativistic fluid the pre-factor
$p(\gamma_\infty-\gamma_0)$ is identical to a commonly used
alternative expression involving the sound speeds of the fluid:
$\rho(u^2_\infty-u^2_0)$~\cite{ll}.  However, this equality is not
satisfied in neutron-star matter.  Consequently it is important to use
the correct expression given in Eq.~(\ref{prefactor}).
\begin{figure}
\includegraphics[width=3.0in]{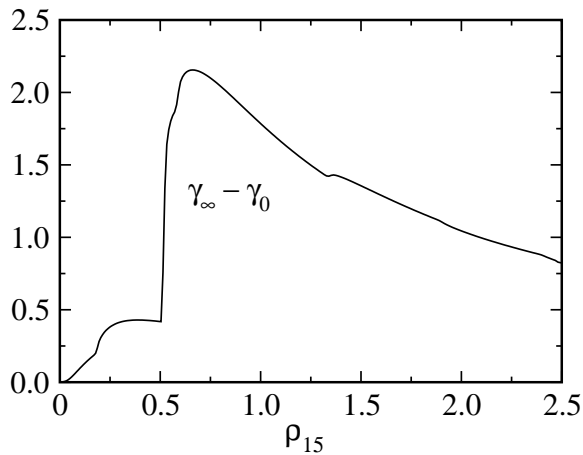}
\caption{Thermodynamic prefactor $\gamma_\infty-\gamma_0$ (the
difference between the ``fast'' and ``slow'' adiabatic indices) that
appears in our expression for the bulk viscosity.\label{f:thermo}}
\end{figure}

\begin{figure}
\vskip 0.4cm
\includegraphics[width=3.0in]{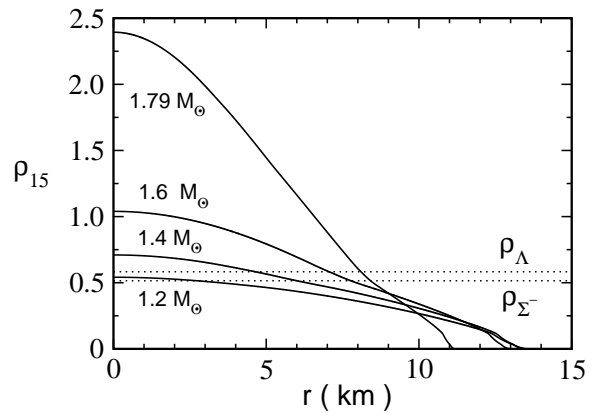}
\caption{Structure of neutron stars having a range of masses using
Glendinning's equation of state: total energy density (in units of
$10^{15}$g/cm$^3$) vs.\ distance from the center of the
star. Threshold densities for $\Sigma^-$ and $\Lambda$ hyperon
formation are also plotted. These equilibrium structures are computed 
using general relativity.\label{f:density}}
\end{figure}

 We have solved the relativistic structure equations for the
non-rotating stellar models based on this equation of state.
Figure~\ref{f:density} illustrates the total energy density as a
function of radius for neutron-star models having a range of
astrophysically relevant masses.  This figure illustrates that these
stars contain large central cores having material at densities that
exceed the $\Sigma^-$ and $\Lambda$ threshold densities.  The fact
that it becomes energetically favorable to create hyperons (or even
free quarks) above some threshold density is not really very
controversial.  However the expressions for the chemical potentials as
functions of the particle densities are not well known, and so the
detailed properties of nuclear matter at the densities where hyperons
are likely to occur is not well determined at this time.  This
uncertainty translates then to an uncertainty about the sizes of the
hyperon containing cores of real neutron stars.  Since the size of
this hyperon core determines the strength of the bulk viscosity
effects which we evaluate here, the implications for the stability
of the $r$-modes are correspondingly uncertain as well.

\subsection{Superfluidity}

\begin{figure}
\includegraphics[width=3.0in]{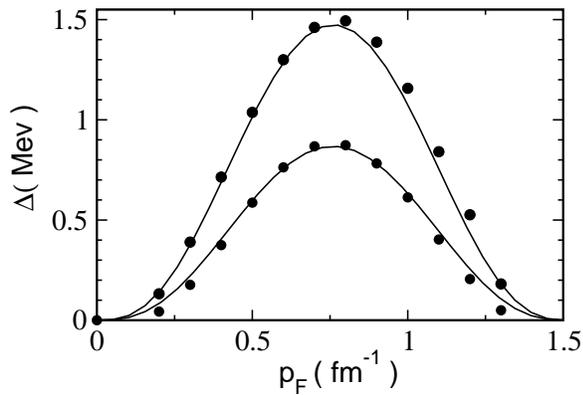}
\caption{Comparison of the zero-temperature supefluid gap function
$\Delta_\Lambda$ as calculated by Balberg and Barnea (dots) with the
empirical analytical fit in Eq.~(\ref{lambda_gap}) 
(curves).  The bottom and top curves correspond to $n_B=0.4$ and 0.8
fm${}^{-3}$ respectively. \label{f:lambda_gaps}}
\end{figure}

Next, we must consider the possibility that the hyperons in
neutron-star matter form Cooper pairs and condense into a superfluid
state at sufficiently low temperatures.  Various calculations are
given in the literature of the $\Lambda$ superfluid gap function
$\Delta_\Lambda$~\cite{balberg,takatsuka}.  The $\Lambda$ gap function
is constrained by the experimental data on the energy levels of double
$\Lambda$ hypernuclei such as ${}^{\,\,10}_{\Lambda\Lambda}$Be and
${}^{\,\,13}_{\Lambda\Lambda}$B~\cite{takatsuka}, however even so it
is probably only known to within a factor of two or three.  In our
numerical analysis of the bulk viscosity timescales discussed in
Sec.~\ref{s:times} we use an analytical fit to the zero-temperature
gap function $\Delta_\Lambda$ as computed by Balberg and
Barnea~\cite{balberg}.  Their calculation produces a gap that depends
on the total baryon density $n_B$ of the nuclear matter and on the
Fermi momentum $p_F$ of the $\Lambda$ itself.  As illustrated in
Fig.~\ref{f:lambda_gaps} we find that the following empirical fit
\begin{eqnarray}
&&\!\!\!\!\!\!\!\!\Delta_\Lambda(p_F,n_B)=5.1 p_F^3(1.52-p_F)^3\nonumber\\
&&\qquad\qquad\times\left[0.77+0.043(6.2n_B-0.88)^2\right],\label{lambda_gap}
\end{eqnarray}
matches their calculated values for the zero-temperature gap energies
fairly well.  In this expression the gap energy is measured in Mev,
$p_F$ is measured in fm${}^{-1}$, and $n_B$ is measured in
fm${}^{-3}$.

\begin{figure}
\vskip 0.8cm
\includegraphics[width=3.0in]{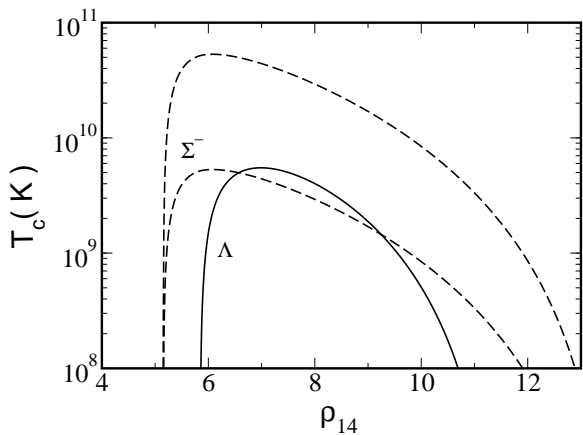}
\caption{Superfluid critical temperatures as functions of the total
energy density (in units of $10^{14}$g/cm${}^3$).  The solid curve is
for $\Lambda$ hyperons, while the dashed curves are for $\Sigma^-$
hyperons using either $\Delta_{\Sigma^-} =\Delta_\Lambda$ (bottom) or
$\Delta_{\Sigma^-}=10\Delta_\Lambda$ (top).\label{f:crit_temp}}
\end{figure}

The $\Sigma^-$ superfluid gaps are not as well determined because
comparable experimental data on double $\Sigma^-$ hypernuclei do not
exist at present.  Calculations by Takatsuka, et al.~\cite{takatsuka}
using several models of the nuclear interaction give values of
$\Delta_{\Sigma^-}$ in the range:
$\Delta_\Lambda\lesssim\Delta_{\Sigma^-} \lesssim10\Delta_\Lambda$.
We perform two sets of calculations based on the extremes of this
range.  We either set $\Delta_{\Sigma^-} =\Delta_\Lambda$ or
$\Delta_{\Sigma^-} =10\Delta_\Lambda$.  By equality here we mean
that the dependence of $\Delta_{\Sigma^-}$ on $n_B$ and $p_F$ (up to
the overall factor of 10) is given by Eq.~(\ref{lambda_gap}).  Using
these energy gaps, and the Glendenning equation of state, we can
evaluate then the density dependence of the superfluid gap
functions.  We illustrate these in the form of superfluid
critical temperatures $T_c$ [which are related to the zero-temperature
gap functions by $kT_c=0.57\Delta(0)$] in Fig.~\ref{f:crit_temp}.

\begin{figure}
\vskip 0.4cm
\includegraphics[width=3.0in]{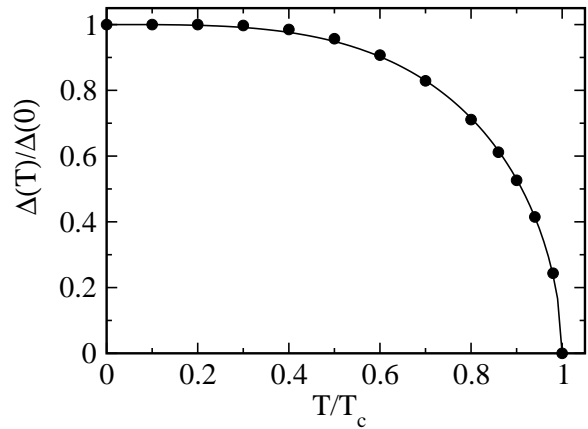}
\caption{Temperature dependence of the superfluid gap: exact values
(points) compared to the empirical analytical fit (curve) given in
Eq.~(\ref{gap_temp}).  The critical temperature $T_c$ is related to
the zero-temparature gap energy by $kT_c=0.57\Delta(0)$.\label{f:gap_temp}}
\end{figure}

The superfluid gap depends not only on the density of the superfluid
material, as discussed above, but also on its temperature.  This
temperature dependence will be needed to determine the temperature
dependence of the hyperon bulk viscosity below.  The standard BCS
model calculation~\cite{muhlschlegel} of the temperature dependence of
the gap is illustrated in Fig.~\ref{f:gap_temp}.  This figure compares
the results of the exact calculation with a simple empirical fit to
these data:
\begin{equation}
\Delta(T)=\Delta(0)\left[1-\left({T\over T_c}\right)^{3.4}\right]^{0.53}.
\label{gap_temp}
\end{equation}
Since this fit is quite good, we use it whenever the temperature
dependence of the gap is needed.

\section{Bulk viscosity}
\label{s:bulk}

Bulk viscosity is the dissipative process in which the macroscopic
compression (or expansion) of a fluid element is converted to heat.
The formalism for calculating the bulk viscosity coefficient in terms
of the relaxation times of the microscopic processes which effect the
conversion is well-known---see, e.g., Landau and
Lifschitz~\cite{ll}. However, such calculations are generally
performed for an ordinary fluid like air, in which the microscopic
processes (typically involving the transfer of energy between
rotational and vibrational degrees of freedom of the molecules) can be
taken to be independent. In neutron-star matter there are several
relevant microscopic processes (involving weak interactions between
the various particle species), but they are related by constraints
(such as conservation of baryon number)---meaning that we cannot
simply use the standard formulas for either a single process or
multiple processes.  Further, neutron-star matter is composed of
some particles having relativistic energies.  The standard
expressions for bulk viscosity are not correct for such materials.  In
this section we present a modified and expanded derivation of the
equations of bulk viscosity appropriate for neutron-star matter.

Bulk viscosity is due to an instantaneous difference between the total
physical pressure $p$ of a fluid element and the thermodynamic
pressure $\tilde p$.  The thermodynamic pressure is determined only by
the equation of state for a fluid element of given particle number and
entropy densities.  It is the value toward which the microscopic
processes are driving the physical pressure at any given time.  The
coefficient of bulk viscosity $\zeta$ defines the proportionality of
this pressure difference to the macroscopic expansion of the fluid:
\begin{equation}
\label{zdef}
p - \tilde p = -\zeta \vec \nabla \cdot \vec v,
\end{equation}
where $\vec v$ is the velocity of the fluid element.

Consider now a fluid state that is an infinitesimal perturbation of a
time-independent equilibrium state.  Let $p_0$ and $n_0$ denote the
pressure and number density (functions only of position) that describe
this equilibrium state.  To calculate $\zeta$ in terms of the
microscopic reaction rates that drive the system toward equilibrium,
we re-express both sides of the Eq.~(\ref{zdef}) in terms of the
Lagrangian perturbation of the particle number density $\Delta n
\equiv n - n_0$. Using the particle conservation equation (and keeping only
terms linear in the deviation away from equilibrium), we express the
right side of Eq.~(\ref{zdef}) as
\begin{equation} 
-\zeta \vec \nabla \cdot \delta \vec v = -i\hat\omega \zeta \Delta n/n,
\label{rhs}
\end{equation}
where $\delta \vec v$ is the Eulerian velocity perturbation, and we assume
that the perturbation has time dependence $e^{-i\hat\omega t}$ in the
comoving frame of the fluid.

In order to analyze the left side of Eq.~(\ref{zdef}), we examine
a fluid variable $x$ that characterizes the microscopic process
which produces bulk viscosity.  For small departures from equilibrium,
the value of $x$ in the physical fluid state relaxes toward its value
in thermodynamic equilibrium by
\begin{equation}
\label{tdef}
\partial_t x + \vec v\cdot\vec\nabla x = - (x-\tilde x) /\tau,
\end{equation}
where $\tau$ is defined as the relaxation time for this process.  We are
interested in nearly equilibrium fluid states in which the physical values
of the state variable $x$ (and hence the thermodynamic state $\tilde x$)
oscillate about the background equilibrium, so that
\begin{eqnarray}
(\partial_t+\vec v\cdot\vec\nabla)(x - {x}_0) &=& -i\hat\omega(x-x_0),
\\
(\partial_t+\vec v\cdot\vec\nabla)(\tilde x - {x}_0) &=& -i\hat\omega(
\tilde x-x_0).
\end{eqnarray}
In such a state it is straightforward to verify that
\begin{eqnarray}
\label{amp}
x -{x}_0 = {x-\tilde x \over i\hat\omega\tau} 
= {\tilde x-{x}_0 \over 1-i\hat\omega\tau}.
\end{eqnarray}

Now consider how the fluid variable $\tilde x$ changes as the particle
number density of the state is varied slowly from one equilibrium
state to another:
\begin{equation}
\tilde x - x_0 = {d \tilde x\over d n}(\tilde n -n_0)
= {d \tilde x\over d n}\Delta n
\label{deltax}
\end{equation}
(since by definition $\tilde n=n$).  It follows then that the
difference between $\tilde p$ and $p_0$ is given by
\begin{equation}
\tilde p-p_0=
\left[\left({\partial p\over \partial n}\right)_x
+\left({\partial p\over \partial x}\right)_n
{d \tilde x\over d n}\right]\Delta n.
\label{deltap1}
\end{equation}
A similar argument (now using Eq.~\ref{amp} to relate $x-x_0$ and $\tilde
x-x_0$) gives the following expression for $p-p_0$:
\begin{equation}
p-p_0=\left[\left({\partial p\over \partial n}\right)_x
+{1\over 1-i\hat\omega\tau}\left({\partial p\over \partial x}\right)_n
{d \tilde x\over dn}\right]\Delta n.
\label{deltap2}
\end{equation}
Combining Eqs.~(\ref{deltap1}) and (\ref{deltap2}) gives us an
expression for the difference in the pressure $p-\tilde p$ that appears on
the left side of Eq.~(\ref{zdef}):
\begin{equation}
p-\tilde p ={i\hat\omega\tau\over 1-i\hat\omega\tau}
\left({\partial p\over \partial x}\right)_n
{d \tilde x \over d n}\Delta n.
\end{equation}
Then equating this expression for the left side of Eq.~(\ref{zdef}) with the
expression for the right side from Eq.~(\ref{rhs}), we find the desired
formula for the bulk viscosity:
\begin{equation}
\zeta={-n\tau\over 1-i\hat\omega\tau}
\left({\partial p\over\partial x}\right)_n
{d \tilde x\over d n}.
\label{zeta}
\end{equation}

Finally it is convenient to re-express the thermodynamic derivatives that
appear in Eq.~(\ref{zeta}) in terms of the more familiar
\begin{equation}
\gamma_\infty = {n\over p}\left({\partial p\over \partial n}\right)_x,
\label{gammai}
\end{equation}
the ``infinite'' frequency adiabatic index, and
\begin{equation}
\gamma_0= {n\over p}\left[\left({\partial p\over \partial n}\right)_x
+\left({\partial p\over\partial x}\right)_n
{d \tilde x\over d n}\right],\label{gamma0}
\end{equation}
the ``zero'' frequency adiabatic index.  In terms of these quantities,
then the bulk viscosity may be written in the form:
\begin{equation}
\zeta = {p(\gamma_\infty - \gamma_0)\tau \over 1-i\hat\omega\tau}.
\label{zeta2} 
\end{equation}
For a fluid composed of particles with non-relativistic energies this
expression is equivalent to the conventional one~\cite{ll} written in
terms of the sound speed $u$, since $p\gamma = \rho u^2$ for such
fluids.  However for a fluid containing particles with relativistic
energies, the conventional form is wrong and Eq.~(\ref{zeta2}) is the
appropriate form to use.  Note that this new form of the bulk
viscosity is needed to describe any fluid containing relativistic
internal particle energies, even if the bulk motion of the fluid
itself has only non-relativistic velocities which are well approximated
by Newtonian hydrodynamics.  We also see that for a fixed-frequency
perturbation, the greatest bulk viscosity comes from processes with
relaxation times $\tau \approx 1/\hat\omega$.  The importance of
this fact will become apparent as we examine how the relaxation time
$\tau$ varies inside a neutron star.

The standard approach~\cite{ll} to treating multiple reactions is to
repeat the preceding derivation for multiple degrees of freedom $x_i$
and relaxation times $\tau_i$. However, this only works if (as in air)
the $x_i$ can be chosen to be independent. This is not possible in
neutron-star matter, since the degrees of freedom (e.g.,
concentrations of various baryons) are related to each other even out
of thermodynamic equilibrium by constraints such as conservation of
baryon number.  The reactions of interest to us here are the
non-leptonic weak interactions
\begin{eqnarray}
n + n &\leftrightarrow& p^+ + \Sigma^-,
\nonumber\\
n + p^+ &\leftrightarrow& p^+ + \Lambda.\nonumber
\end{eqnarray}
Given the microscopic reaction rates for these processes (which are
calculated in Sec.~\ref{s:rates}), we can express all of the perturbed
quantities in terms of a single one. Since all the hyperon reactions
that contribute to bulk viscosity involve neutrons, we choose as our
primary variable the number density of neutrons $n_n$.

Let $x_n = n_n/n_B$ be the fraction of baryons in a given fluid element that
are neutrons. This variable changes only by internal reactions, not directly
by changing the volume of the fluid element, and so we can write
\begin{equation}
(\partial_t +\vec v\cdot\vec\nabla)x_n = - (x_n-\tilde x_{n}) /\tau 
= -\Gamma_n/n_B.
\end{equation}
Here $\Gamma_n$ is the production rate of neutrons per unit volume,
which is proportional to the overall chemical potential imbalance
$\delta\mu\equiv \mu-\tilde\mu$ (see below). (We normalize $n_n$ and
$\Gamma_n$ to the baryon number density $n_B$ to remove the
oscillating time dependence of the volume of the fluid element.)
Assuming the reactions described in Sec.~\ref{s:rates}, the relaxation
time is then given by
\begin{equation}
{1\over\tau} = {\Gamma_\Lambda + 2\Gamma_\Sigma \over \delta\mu} {\delta\mu
\over n_B\delta x_n},\label{tau}
\end{equation}
where $\delta x_n\equiv x_n-\tilde x_n$.

We obtain $\delta\mu /\delta x_n$ from the following constraints:
\begin{eqnarray}
0 &=& \delta x_n + \delta x_\Lambda + \delta x_p + \delta x_\Sigma,
\label{const1}\\
0 &=& \delta x_p - \delta x_\Sigma,
\\
0 &=& \beta_n \delta x_n + \beta_\Lambda \delta x_\Lambda + \beta_p \delta
x_p + \beta_\Sigma \delta x_\Sigma,\label{const3}
\end{eqnarray}
where $x_i$ are the number densities of baryon species normalized to
the total number density of baryons and the $\beta_i$ are defined
below. The first constraint is conservation of baryon number, obeyed
by all reactions.  (We note that since $\delta n_B = n_B-\tilde
n_B\equiv0$ that $\delta x_i = \delta n_i/n_B$.)  The second
constraint is related to conservation of electric charge, but is
stricter: We assume that all leptonic (Urca) reaction rates are much
smaller than those which produce hyperon bulk viscosity
and so protons are only produced in reactions that produce a
$\Sigma^-$. The third constraint is that the non-leptonic reaction
\begin{equation}
\label{strong}
n + \Lambda \leftrightarrow p^+ + \Sigma^-
\end{equation}
proceeds much faster than the weak interactions which produce hyperon bulk
viscosity since it is mediated by the strong nuclear interaction.
Here we use the shorthand
\begin{eqnarray}
\alpha_{ij} &=& \left( \partial \mu_i \over \partial n_j \right)_{n_k, 
{\scriptscriptstyle k\ne j}}
\\
\beta_i &=& \alpha_{ni} + \alpha_{\Lambda i} - \alpha_{pi} - \alpha_{\Sigma
i}.
\end{eqnarray}
Equilibrium with respect to reaction~(\ref{strong}) ensures that both
processes described in Sec.~\ref{s:rates} have the same chemical potential
imbalance,
\begin{equation}
\delta\mu \equiv \delta\mu_n - \delta\mu_\Lambda = 2\delta\mu_n - \delta\mu_p -
\delta\mu_\Sigma.
\end{equation}
It is straightforward then to express $\delta \mu$ in terms of the
$\delta x_i$, and then to eliminate all but $\delta x_n$ using the 
constraints Eqs.~(\ref{const1})--(\ref{const3}).   The result,
\begin{eqnarray}
&&{\delta \mu\over n_B\delta x_n} = \alpha_{nn} +
{(\beta_n-\beta_\Lambda)(\alpha_{np}-\alpha_{\Lambda p}+\alpha_{n\Sigma}
-\alpha_{\Lambda\Sigma})\over 2\beta_\Lambda-\beta_p-\beta_\Sigma}\nonumber\\
&&\,\,\qquad-\alpha_{\Lambda n}
-{(2\beta_n-\beta_p-\beta_\Sigma)(\alpha_{n\Lambda}-\alpha_{\Lambda\Lambda})
\over 2\beta_\Lambda -\beta_p-\beta_\Sigma},\label{delta1}
\end{eqnarray}
then determines via Eq.~(\ref{tau}) the relaxation time that appears in the 
bulk viscosity formula Eq.~(\ref{zeta2}).  For a certain range of densities
there are $\Sigma^-$ hyperons present in Glendinning's equation of state,
but no $\Lambda$.  In that case the variable $\delta x_\Lambda$ remains
zero, and the constraint Eq.~(\ref{const3}) is no longer enforced.  In this
case the chemical potential imbalance can still be expressed in terms of
$\delta x_n$ with the somewhat simpler result:
\begin{eqnarray}
{2\delta \mu\over n_B\delta x_n} &=& 
4\alpha_{nn}-2(\alpha_{pn}+\alpha_{\Sigma n}+\alpha_{np}+\alpha_{n\Sigma})
\nonumber\\
&&+\alpha_{pp}+\alpha_{\Sigma p}
+\alpha_{p\Sigma}+\alpha_{\Sigma\Sigma}.\label{delta2}
\end{eqnarray}

The equation of state of neutron-star matter is generally written in a
form which gives the thermodynamic variables as functions of the
various particle species present, e.g. the pressure would be specified
as $p=p(n_i)$.  In order to evaluate the thermodynamic derivatives
$(\partial p/\partial n)_x$ and $(\partial p/\partial x)_n$ that are
needed in Eq.~(\ref{zeta}), we note that $n_n=n_B x_n$ for our choice of
$x$.  Thus the partial derivative needed in Eq.~(\ref{zeta}) is given
by $(\partial p/\partial x)_n=n_B\partial p/\partial n_n$.  Similarly,
if needed, $(\partial p/\partial n)_x=\sum_i x_i\partial p/\partial
n_i$.  The derivative $d\tilde x/d n$ that also appears
in Eq.~(\ref{zeta}) is determined by constructing a sequence of
complete equilibrium models (e.g. by imposing all of the necessary
$\beta$-equilibrium constraints, etc.) for different total baryon
number densities $n_B$.  This complete model of the equilibrium states
will include the functions $\tilde x_n(n_B)$, from which the derivative
$d \tilde x_n/d n_B$ is easily computed.

Formation of a superfluid (of a given particle species) is marked by
the formation of Cooper pairs and collapse of the pairs into a
Bose-Einstein condensate. It is the unpaired particles that we are
concerned with, since they are the ones that can participate in the
bulk-viscosity generating reactions. The free-particle states within
the pair-binding energy $\Delta$ of the Fermi surface are depleted.
As a result all phase-space factors (and effectively the reaction
rates) are decreased by roughly a factor $e^{-\Delta/kT}$.  The effect
of superfluidity can be included in our ordinary-fluid bulk viscosity
calculation then simply by making the substitution
\begin{equation}
\Gamma \to e^{-\Delta/kT} \Gamma.
\end{equation}
Thus when superfluidity is taken into account the equation
for the relaxation time, Eq.~(\ref{tau}) becomes:
\begin{equation}
{1\over\tau} = \left({\Gamma_\Lambda\over \delta\mu }e^{-\Delta_\Lambda/kT} 
+ {2\Gamma_\Sigma \over \delta\mu }e^{-\Delta_\Sigma/kT}\right) {\delta\mu
\over n_B\delta x_n}.\label{tau2}
\end{equation}

\section{Microscopic reaction rates}
\label{s:rates}

Since the $\Sigma^-$ and $\Lambda$ hyperons form at the lowest threshold
densities, we are most interested in the nonleptonic reactions forming them
from neutrons.  Following Jones' earlier work~\cite{jones71,jones01} we calculate rates for two reactions,
\begin{eqnarray}
\label{reacS}
n + n &\leftrightarrow& p^+ + \Sigma^-,
\\
\label{reacL}
n + p^+ &\leftrightarrow& p^+ + \Lambda,
\end{eqnarray}
as tree-level Feynman diagrams involving the exchange of a $W$ boson.
In his most recent paper~\cite{jones2}, Jones treats the reaction
\begin{equation}
\label{jones2}
n + n \leftrightarrow n + \Lambda,
\end{equation}
which is generally the dominant channel for $\Lambda$ production in
laboratory experiments on hypernuclei.  This process has no simple
$W$-exchange contribution. Several other processes contribute, but at
present the rate cannot be well predicted from theory.  Some of these
processes surely operate in reactions~(\ref{reacS}) and~(\ref{reacL})
and modify the rates, perhaps significantly. However, our simple
calculations should provide a reasonable lower limit on the rates and
thus an upper limit on the bulk viscosity.

\subsection{Single reactions}

We calculate reaction rates using the standard techniques of time-dependent
perturbation theory in relativistic quantum mechanics.  We use the
conventions of Griffiths~\cite{grif}: spinors are normalized to
$\bar{u}u=2m$, the fifth Dirac matrix is $\gamma^5 = i\gamma^0 \gamma^1
\gamma^2 \gamma^3$, the metric has negative trace, and $\hbar=1$.  For a
single reaction~(\ref{reacS}) or~(\ref{reacL}) between particles with
4-momenta $p_i$, the differential reaction rate (number per unit volume per
unit time) is
\begin{equation}
\label{dGam}
d\Gamma = |{\cal M}|^2 (2\pi)^4 \delta^{(4)} (p_1 + p_2 - p_3 - p_4)
S\prod_{i=1}^4 {d^3\bm{p}_i \over (2\pi)^3 2\epsilon_i} ,
\end{equation}
where $|{\cal M}|^2$ is the spinor matrix element (squared and summed over
spin states), boldface $\bm{p}_i$ are 3-momenta, and $\epsilon_i$ are particle
energies. The statistical factor $S$, which compensates for overcounting
momentum states of indistinguishable particles, is $1/2$ for
reaction~(\ref{reacS}) and $1$ for reaction~(\ref{reacL}).

First, consider the $\Lambda$ reaction~(\ref{reacL}), which we represent by
a single tree-level diagram. Labeling the particles 1 (neutron), 2 (ingoing
proton), 3 (outgoing proton), 4 ($\Lambda$ hyperon), we obtain the matrix
element (for a single set of spin states)
\begin{eqnarray}
\label{M_L}
{\cal M}_\Lambda &=& {G_F \over 2\sqrt{2}} \sin 2\theta_C \left[
\bar{u}(p_3) \gamma^\mu \left(1 + \anp\gamma^5\right) u(p_1) \bar{u}(p_4)
\right. \nonumber
\\
&& \times \left. \gamma_\mu \left(1 + \apL\gamma^5\right) u(p_2) \right] .
\end{eqnarray}
Here $G_F$ is the Fermi coupling constant and $\theta_C$ is the
Cabibbo weak mixing angle. The quantities $\anp$, $\anS$, and $\apL$
are axial-vector couplings (normalized to the vector coupling) of the
weak interaction, whose deviation from $-1$ represents the partial
nonconservation of the axial current. (These quantities are often
written $g_A$ or $G_A/G_V$ in the particle and nuclear physics
literature. We add a label to keep track of which nucleon-hyperon line
is which.) We use the values $G_F = 1.166\times10^{-11}$MeV$^{-2}$ and
$\sin\theta_C = 0.222$ from the Particle Data Group~\cite{pdg}. The
axial-vector couplings change with varying momentum transfer and
density of the medium in a way that reflects the internal
(strong-interaction) structure of the baryons and is therefore
difficult to calculate. We use the values $\anp = -1.27$,
$\apL = -0.72$, and $\anS = 0.34$ measured in $\beta$-decay of baryons
at rest~\cite{pdg}. There are theoretical reasons to believe that all
axial-vector couplings tend to their asymptotically free values of
$-1$ in a dense medium~\cite{carter-prakash}. We provide dissipation numbers
in later sections both for laboratory values and for asymptotic values of
the couplings to give an estimate of the uncertainty in this calculation.

\begin{widetext}
To obtain the net reaction rate $\Gamma$ we sum $|{\cal M}|^2$ over all
possible initial and final spin states. This is done in the standard way by
tracing over outer products of spinors:
\begin{eqnarray}
\label{msqL}
|{\cal M}_\Lambda|^2 &=& 4G_F^2 \sin^22\theta_C \left\{ 2 m_nm_p^2m_\Lambda
\left(1-\anp^2\right) \left(1-\apL^2\right) - m_nm_p p_2 \cdot p_4
\left(1-\anp^2\right) \left(1+\apL^2\right) \right. \nonumber
\\
&& - m_pm_\Lambda p_1 \cdot p_3 \left(1+\anp^2\right) \left(1-\apL^2\right)
+ p_1 \cdot p_2\, p_3 \cdot p_4 \left[ \left(1+\anp^2\right)
\left(1+\apL^2\right) + 4\anp\apL \right] \nonumber
\\
&& \left.  + p_1 \cdot p_4\, p_2 \cdot p_3 \left[ \left(1+\anp^2\right)
\left(1+\apL^2\right) - 4\anp\apL \right] \right\},
\end{eqnarray}
which in the low-momentum limit used by Jones~\cite{jones2} reduces to
\begin{equation}
|{\cal M}_\Lambda|^2 = 8G_F^2 \sin^22\theta_C m_n m_p^2 m_\Lambda \left(1 +
3\anp^2\apL^2\right).
\end{equation}
Note that this limit corresponds to $G_\Lambda=0.40G_F$ in the notation of
Jones~\cite{jones2}, who obtains $G_\Lambda=1.29G_F$. The factor of three
discrepancy is within the uncertainties of modern nuclear-matter physics, as
we see by e.g.\ taking the asymptotic values of the axial-vector couplings.

Reaction~(\ref{reacS}) is treated similarly, with particle labels 1 and 2
(neutrons), 3 (proton), and 4 ($\Sigma^-$ hyperon). Antisymmetrizing with
respect to the two indistinguishable neutrons, the matrix element is
\begin{eqnarray}
\!\!\!\!\!\!{\cal M}_\Sigma &=& {G_F\over 2\sqrt{2}}\sin 2\theta_C
\bar{u}(p_3) \gamma^\mu \left(1 + \anp\gamma^5\right) 
\left[u(p_1) \bar{u}(p_4) \gamma_\mu \left(1 + \anS\gamma^5 \right) u(p_2)
- u(p_2) \bar{u}(p_4) \gamma_\mu \left(1 + \anS\gamma^5 \right) u(p_1)\right].
\end{eqnarray}
The squared sum over spins is given by
\begin{eqnarray}
\label{msqS}\!\!\!\!\!\!\!\!\!\!
|{\cal M}_\Sigma|^2 &=& 4G_F^2 \sin^22\theta_C \left\{ 6m_n^2m_pm_\Sigma
\left(1-\anp^2\right) \left(1-\anS^2\right) - m_pm_\Sigma p_1 \cdot p_2
\left(1-\anp^2\right) \left(1-\anS^2\right) \right. \nonumber
\\
&& - 2m_nm_\Sigma p_1 \cdot p_3 \left(1+\anp^2\right) \left(1-\anS^2\right)
- 2m_nm_p p_1 \cdot p_4 \left(1-\anp^2\right) \left(1+\anS^2\right)
\nonumber
\\
&& - 2m_nm_\Sigma p_2 \cdot p_3 \left(1+\anp^2\right) \left(1-\anS^2\right)
- 2m_nm_p p_2 \cdot p_4 \left(1-\anp^2\right) \left(1+\anS^2\right)
\nonumber
\\
&& - m_n^2 p_3 \cdot p_4 \left[ \left(1+\anp^2\right) \left(1+\anS^2\right)
- 4\anp\anS \right] 
+ 4p_1 \cdot p_2\, p_3 \cdot p_4 \left[ \left(1+\anp^2\right)
\left(1+\anS^2\right) + 4\anp\anS \right] \nonumber
\\
&& + p_1 \cdot p_3\, p_2 \cdot p_4 \left[ \left(1+\anp^2\right)
\left(1+\anS^2\right) - 4\anp\anS \right]
+ p_1 \cdot p_4\, p_2 \cdot p_3 \left[ \left(1+\anp^2\right)
\left(1+\anS^2\right) - 4\anp\anS \right] \left. \right\} ,
\end{eqnarray}
\end{widetext}
which in the low-momentum limit reduces to
\begin{equation}
\label{low}
|{\cal M}_\Sigma|^2 = 8G_F^2 \sin^22\theta_C m_n^2m_pm_\Sigma \left(1 +
3\anp\anS\right)^2.
\end{equation}
This expression agrees with Eq.~(7) of Jones' old paper~\cite{jones71},
allowing for our different placement of the statistical overcounting factor
and different normalization of the spinors (he uses $\bar{u}u=1$). However,
we find that for this reaction the low-momentum limit is a very poor
approximation to the full result: the collision integral computed in the
next subsection can be more than an order of magnitude higher than
Eq.~(\ref{low}) would suggest. This effectively brings down the coefficient
of bulk viscosity $\zeta$ by a factor of about 5 at densities
5--6$\times10^{14}$~g/cm$^3$, where there are no $\Lambda$
hyperons~\cite{haensel}.

\subsection{Collision integrals}

Reactions~(\ref{reacS}) and~(\ref{reacL}) can be regarded as scattering
processes wherein the scattered particles change identity. Therefore we can
use existing results on collision integrals in the literature on
superfluidity (e.g.~\cite{pn}) with some slight modifications.  We now
integrate the differential reaction rate~(\ref{dGam}) over momentum space to
obtain the total rate
\begin{eqnarray}
\label{collision}
\Gamma &=& 
{S\over 4096\pi^8} \int \prod_{i=1}^4 {d^3 \bm{p}_i \over \epsilon_i}
|{\cal M}|^2
\delta^{(3)} (\bm{p}_1 + \bm{p}_2 - \bm{p}_3 - \bm{p}_4) \nonumber
\\
&& \times   F(\epsilon_i)\delta (\epsilon_1+\epsilon_2-\epsilon_3-\epsilon_4).
\end{eqnarray}
(From here on we use italic $p_i$ to denote the absolute value of the
3-momenta, $p_i = \sqrt{\bm{p}_i \cdot \bm{p}_i}$.) The Pauli blocking
factor,
\begin{equation}
F(\epsilon_i)\! =\! f_1f_2(1-f_3)(1-f_4) - (1-f_1)(1-f_2)f_3f_4,
\end{equation}
where
\begin{equation}
f_i = 1/\{1 + \exp[(\epsilon_i-\mu_i)/kT]\} ,
\end{equation}
accounts for the degeneracy of the reactant particles and restricts the
available phase space to those particles within roughly $kT$ of their
Fermi energies.

In the case where all particles are degenerate (which is true except for a
very small region just above the threshold density for each hyperon
species), the collision integral separates into angular and energetic parts.
The energetic part can be written in the limit $\delta\mu \ll kT$ as
\begin{eqnarray}
\label{energy}
&&\!\!\!\!\!\!\!\!\!\!\!\!\!
\int \prod_i d\epsilon_i F(\epsilon_i) \delta(\epsilon_1+\epsilon_2
-\epsilon_3-\epsilon_4) \nonumber
\\
&&\quad = (kT)^2 \delta\mu
\int_{-\infty}^{+\infty} {y^2 dy \over \left(e^{y}-1\right)
\left(1-e^{-y}\right)},\quad
\end{eqnarray}
where the latter integral has the value $2\pi^2/3$.

\begin{figure}
\includegraphics[width=2.5in]{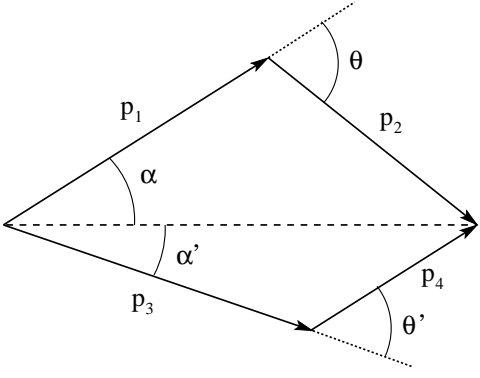}
\vskip 0.3cm
\caption{Definition of angles used in the collision integral. The plane
containing $\bm{p}_3$ and $\bm{p}_4$ can be rotated out of the page
around the long-dashed common axis by an angle $\phi'$ for a given
$\bm{p}_1$ and $\bm{p}_2$.\label{f:angles}}
\end{figure}

We address the angular integral with the aid of Fig.~\ref{f:angles}.  With
no preferred direction (e.g., strong magnetic field), the $\bm{p}_1$
volume element is $4\pi p_1^2 dp_1$ and the $\bm{p}_2$ volume element is
$p_2^2 dp_2 \sin\theta d\theta\, d\phi$, where $\theta$ and $\phi$ are the
polar and azimuthal angles of $\bm{p}_2$ with respect to $\bm{p}_1$.
Conservation of momentum demands that $\bm{p}_1 + \bm{p}_2 = \bm{p}_3
+ \bm{p}_4$, determining a common axis. Let $\alpha$ and $\alpha'$ be the
angles of $\bm{p}_1$ and $\bm{p}_3$ with that axis. The volume element
of $\bm{p}_3$ is then $p_3 \sin\alpha' d\phi'$, where $\phi'$ is the angle
between $\bm{p}_1 \times \bm{p}_2$ and $\bm{p}_3 \times \bm{p}_4$,
multiplied by the area element in the plane containing $\bm{p}_3$ and
$\bm{p}_4$. To separate the energetic and angular integrals, it is
convenient to write this area element as $dp_3 dp_4 /\sin\theta'$, where
$\theta'$ is the angle between $\bm{p}_3$ and $\bm{p}_4$. Before leaving
Fig.~\ref{f:angles}, note the following useful identities:
\begin{eqnarray}
\!\!\!\!\!\!\!\!\!\!
p_1^2 + p_2^2 + 2p_1p_2 \cos\theta &=& p_3^2 + p_4^2 + 2p_3p_4 \cos\theta',
\\
p_1 \sin\alpha &=& p_2 \sin(\theta-\alpha) ,
\\
p_3 \sin\alpha' &=& p_4 \sin(\theta'-\alpha') .
\end{eqnarray}
We use the differential of the first identity (with $p_i$ held constant) to
write our final result for the volume element as
\begin{eqnarray}
\delta^{(3)} (\bm{p}_1 + \bm{p}_2 - \bm{p}_3 - \bm{p}_4)
\prod_{i=1}^4 {d^3\bm{p}_i \over \epsilon_i}\qquad\qquad&& \nonumber
\\
= 4\pi p_3 \sin\alpha' d\theta' d\phi' d\phi \prod_{i=1}^4 d\epsilon_i .&&
\end{eqnarray}

Although $\phi$ and $\phi'$ are free to range from 0 to $2\pi$, the limits
of integration over $\theta'$ depend on the relations between the momenta
$p_i$, which are constrained to be close to the Fermi momenta. (Therefore
from now on we use $p_i$ to refer to the Fermi momenta.) We integrate
$\theta'$ over the full range $0$ to $\pi$, which is allowed by momentum
conservation for
\begin{eqnarray}
\label{triangles}
p_n \ge p_p \ge (p_\Lambda, p_\Sigma),\quad
p_n - p_p \le p_p - p_\Lambda.
\end{eqnarray}
In our equation of state these Fermi momentum criteria are satisfied for
$\rho < 1.0\times10^{15}$g/cm$^3$ (see Fig.~\ref{f:p_fermi}). It turns out
that almost all of the dissipation takes place below this density, and much
above this density the baryons are too closely packed to maintain their
separate identities anyway. Therefore for our purposes it is sufficient to
treat only the case~(\ref{triangles}).
 
Now consider the case that $|{\cal M}|^2$ does not depend on any of the
angles, such as in the limit $p_i/m_i \ll1$. We can integrate trivially over
all the angles, with the exception of
\begin{equation}
\int_0^\pi p_3 \sin\alpha' d\theta' = 2p_4,
\end{equation}
where $p_4$ is the smallest (i.e.\ hyperon) Fermi momentum involved. This
motivates us to use Eq.~(\ref{energy}) to write
\begin{equation}
\label{rate}
\Gamma = {S \over 192\pi^3} \langle |{\cal M}|^2 \rangle p_4 (kT)^2
\delta\mu,
\end{equation}
where $\langle |{\cal M}|^2 \rangle$ is the angle-averaged value of $|{\cal
M}|^2$. We write $|{\cal M}|^2$ in terms of the integrals
\begin{eqnarray}
\langle \bm{p}_1 \cdot \bm{p}_2 \rangle &=& {\scriptstyle {1\over6}} 
\left( -3p_1^2 - 3p_2^2 + 3p_3^2 + p_4^2 \right),
\\
\langle \bm{p}_1 \cdot \bm{p}_3 \rangle &=& {\scriptstyle{1\over6}} 
\left( -3p_1^2 +
3p_2^2 - 3p_3^2 + p_4^2 \right),
\\
\langle \bm{p}_1 \cdot \bm{p}_4 \rangle &=& \langle \bm{p}_2 \cdot
\bm{p}_4 \rangle = \langle \bm{p}_3 \cdot \bm{p}_4 \rangle =
-{\scriptstyle{1\over3}} p_4^2,
\\
\langle \bm{p}_2 \cdot \bm{p}_3 \rangle &=& {\scriptstyle{1\over6}} 
\left( 3p_1^2 -
3p_2^2 - 3p_3^2 + p_4^2 \right),
\\\!\!\!\!\!\!\!\!\!\!\!\!\!
\langle \bm{p}_1 \cdot \bm{p}_2 \, \bm{p}_3 \cdot \bm{p}_4 \rangle
&=& {\scriptstyle{1\over30}} 
p_4^2 \left( 5p_1^2 + 5p_2^2 + 5p_3^2 - p_4^2 \right),
\end{eqnarray}
\begin{widetext}
\begin{eqnarray}
\langle \bm{p}_1 \cdot \bm{p}_3 \, \bm{p}_2 \cdot \bm{p}_4 \rangle
&=&
\langle \bm{p}_1 \cdot \bm{p}_4 \, \bm{p}_2 \cdot \bm{p}_3 \rangle
= {p_4^2 \over 60\left( p_3^2 - p_4^2 \right)}\biggl[ 5p_1^4 -
10p_1^2p_2^2 + 5p_2^2 + 3\left( 5p_3^4 - 6p_3^2p_4^2 + p_4^4 \right)
\biggr].
\end{eqnarray}
(Note that some dot products, e.g.\ $\bm{p}_1 \cdot \bm{p}_3$, depend on
$\phi'$. The average over $\phi'$ is then nontrivial but is easily taken by
symmetry about the common axis.)

The results for the angle averages are
\begin{eqnarray}
\langle |{\cal M}_\Lambda|^2 \rangle&& = {G_F^2 \sin^22\theta_C \over 15}
\left\{ 120\manp\mapL m_nm_p^2m_\Lambda - 20\manp\papL m_nm_p \left(3\ep\eL
- \pL^2\right) \right.\nonumber
\\
&&\left. - 10\panp  
\mapL m_pm_\Lambda \left(6\en\ep - 3p_n^2 + \pL^2\right) + 2\left[
\panp\papL + 4\anp\apL \right] \right.\nonumber\\
&&\left.\times\left[ 5\ep\eL \left( 6\en\ep + 3p_n^2 -
\pL^2 \right) + \pL^2 
\left( 10\en\ep + 5p_n^2 + 10p_p^2 - \pL^2 \right) \right]
+ \left[ \panp\papL - 4\anp\apL \right]\right. \nonumber\\
&&\left.\times\left[ 10\en\eL \left( 6m_p^2 +
3p_n^2 + \pL^2 \right)
+ \pL^2 \left( -20\ep^2 + 15p_p^2 - 3\pL^2 + 5{(p_n^2 -
p_p^2)^2 /( p_p^2 - \pL^2)} \right) \right] \right\}.\label{m2lambda}
\end{eqnarray}
[Note that the denominator in the last term does not diverge while
criterion~(\ref{triangles}) holds.] Similarly, for reaction~(\ref{reacS}),
\begin{eqnarray}
\langle |{\cal M}_\Sigma|^2 \rangle&& = {2\over15} G_F^2 \sin^22\theta_C
\left\{ 180\manp\manS m_n^2m_pm_\Sigma - 40\manp\panS m_nm_p \left( 3\en\eS
- \pS^2 \right) \right.\nonumber
\\
&& - 20\panp\manS m_nm_\Sigma \left( 6\en\ep - 3p_p^2 + \pS^2 \right)
- 5\manp\manS m_pm_\Sigma \left( 6\en^2 + 6p_n^2 - 3p_p^2 -\pS^2 \right)
\nonumber
\\
&& + 4\left[ \panp\panS + 4\anp\anS \right] \left[ 10\en^2 \left( 3\ep\eS +
\pS^2 \right) + 5\ep\eS \left( 6p_n^2 - 3p_p^2 - \pS^2 \right) + \pS^2
\left( 10p_n^2 + 5p_p^2 - \pS^2 \right) \right] \nonumber
\\
&& \left. + 10\left[ \panp\panS - 4\anp\anS \right] \left[ -m_n^2 \left(
3\ep\eS + \pS^2 \right) + \en \left( 6\en\ep\eS - 2\ep\pS^2 - 3\eS p_p^2 +
\eS\pS^2 \right) \right]
\right\}.\label{m2sigma}
\end{eqnarray}
\end{widetext}
These angle averages are inserted into Eq.~(\ref{rate}) to yield the net
reaction rate per unit volume as a function of Fermi momenta.

\section{Relaxation times}
\label{s:times}
The relaxation timescales for the non-leptonic weak interactions,
Eqs.~(\ref{reacS}) and (\ref{reacL}), can now be determined by combining
the microscopic collision rates determined in Sec.~\ref{s:rates} with
the thermodynamic quantities evaluated in Secs.~\ref{s:eos}
and \ref{s:bulk}.  At densities above the threshold for the production
of $\Sigma^-$, but below the $\Lambda$ threshold, the final expression
for the relaxation timescale $\tau$ is,
\begin{equation}
{1\over\tau}={(kT)^2\over 192\pi^3}
{p_\Sigma\langle|{\cal M}_\Sigma|^2\rangle\over e^{\Delta_\Sigma/kT}}
{\delta\mu\over n_B\delta x_n}.
\label{taus}
\end{equation}
Here the collision cross section $|{\cal M}_\Sigma|^2$ 
is evaluated from Eq.~(\ref{m2sigma}).  
The Fermi momenta that appear in this expression are obtained from the
complete description of the equilibrium thermodynamic state as described
in Sec.~\ref{s:eos}.  This description includes the values of the particle
densities (and hence Fermi momenta) of each species as a function of the
total baryon density.  The thermodynamic quantity $\delta\mu/n_B\delta x_n$
is given by Eq.~(\ref{delta2}) in terms of chemical potentials and their
derivatives.  Once the density increases to the point that both $\Sigma^-$
and $\Lambda$ hyperons are present in the equilibrium state, then
the expression for the relaxation timescale becomes,
\begin{equation}
{1\over\tau}={(kT)^2\over 192\pi^3}
\left[{p_\Sigma\langle|{\cal M}_\Sigma|^2\rangle\over e^{\Delta_\Sigma/kT}}
+{p_\Lambda\langle|{\cal M}_\Lambda|^2\rangle\over e^{\Delta_\Lambda/kT}}\right]
{\delta\mu\over n_B\delta x_n}.
\label{tausl}
\end{equation}
Here the collision cross sections are given by Eq.~(\ref{m2lambda})
and (\ref{m2sigma}), and $\delta\mu/n_B\delta x_n$ is given by
Eq.~(\ref{delta1}) in this case.

\begin{figure}
\includegraphics[width=3.0in]{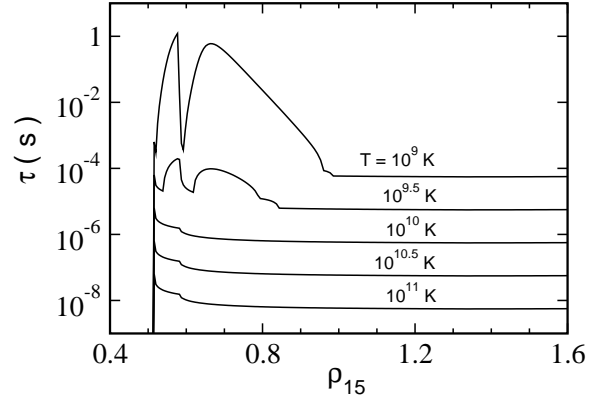}
\caption{Density dependence (in units of $10^{15}$ g/cm${}^3$) of the
relaxation timescale $\tau$ (in units of s) for a range of
temperatures.  These curves were constructed using the assumption
$\Delta_{\Sigma^-}=\Delta_\Lambda$ for the $\Sigma^-$ superfluid gap.
\label{f:tau}}
\end{figure}

\begin{figure}
\includegraphics[width=3.0in]{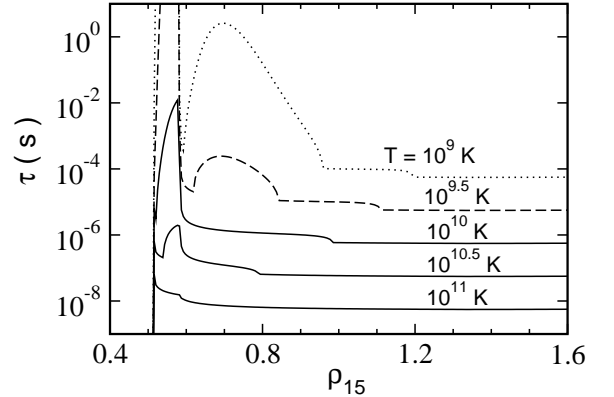}
\caption{Density dependence (in units of $10^{15}$ g/cm${}^3$) of the
relaxation timescale $\tau$ (in units of s) for a range of
temperatures.  These curves were constructed using the assumption
$\Delta_{\Sigma^-}=10\Delta_\Lambda$ for the $\Sigma^-$ superfluid gap.
\label{f:tau10}}
\end{figure}

\begin{figure}
\includegraphics[width=3.0in]{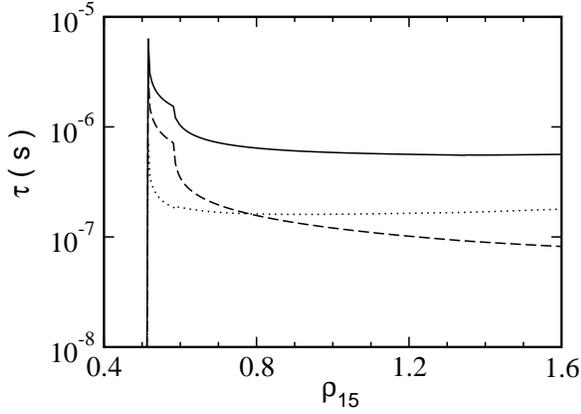}
\caption{Density dependence (in units of $10^{15}$ g/cm${}^3$) of the
relaxation timescale $\tau$ (in units of s) for $T=10^{10}$K.  Solid
curve uses effective masses and the $\beta$-equilibrium values of the
axial-vector couplings. Other cuves explore various alternate
microphysics assumptions: dashed curve uses bare masses, dotted curve
uses $g_{B}=-1$.\label{f:tau_alt}}
\end{figure}

We have evaluated this relaxation timescale for neutron-star matter
using the equation of state described in Sec.~\ref{s:eos}.
Figure~\ref{f:tau} illustrates the density dependence of this
timescale for a range of temperatures.  In the case of
Fig.~\ref{f:tau} we have assumed that the $\Sigma^-$ superfluid gap
function is given by $\Delta_\Sigma=\Delta_\Lambda$, while in
Fig.~\ref{f:tau10} we assume $\Delta_\Sigma=10\Delta_\Lambda$.  The
only significant difference between these two cases comes about in the
density range where there exist $\Sigma^-$ but not $\Lambda$.  In that
range the timescale is significantly increased in the
$\Delta_\Sigma=10\Delta_\Lambda$ case by the stronger superfluid
effects.  All of the curves in these two figures were evaluated using
the ``standard'' $\beta$-decay values of the axial-vector coupling
constants: $g_{np}=-1.27$, $g_{p\Lambda}=-0.72$, and
$g_{n\Sigma}=0.34$~\cite{pdg} and the effective masses of all of the
baryons when evaluating the scattering cross sections.  There are
theoretical arguments that suggest the coupling constants should
approach the values, $g_{np}=g_{p\Lambda}=g_{n\Sigma}=-1$ in a dense
medium~\cite{carter-prakash}.  We illustrate the impact this might
have on these timescales in Fig.~\ref{f:tau_alt} (dotted curve).  We
also illustrate in this figure (dashed curve) the effect of using the
bare masses of the various baryons when computing the scattering cross
sections.  We see that the overall effect of these changes is to make
the timescales shorter (by up to an order of magnitude).  This
tends to decrease the bulk viscosity by a similar factor, until the
temperature drops below the superfluid critical values.

\begin{figure}
\includegraphics[width=3.0in]{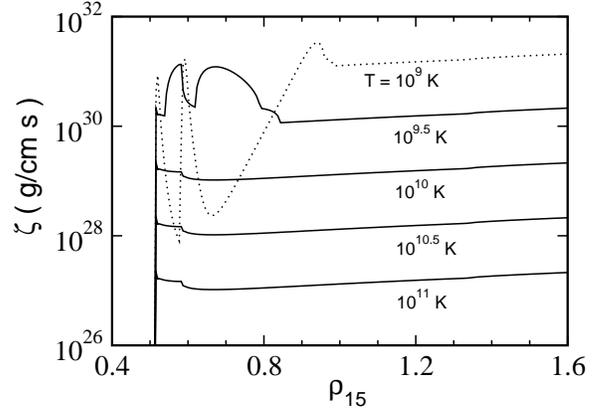}
\caption{Density dependence (in units of $10^{15}$ g/cm${}^3$) of
the hyperon bulk viscosity (in units of g/cm s) for a range
of temperatures.  These curves were constructed using the assumption
$\Delta_{\Sigma^-}=\Delta_\Lambda$ for the $\Sigma^-$ superfluid gap.
\label{f:zeta}}
\end{figure}

Finally, we are in a position now to evaluate the bulk viscosity itself.
The real part of the bulk viscosity, the part that is responsible for
damping the modes of neutron stars, is given by the expression
\begin{equation}
{\rm Re}\,\,\zeta={p(\gamma_\infty-\gamma_0)\tau\over
1+(\hat\omega \tau)^2}.\label{realzeta}
\end{equation}
Figure~\ref{f:zeta} illustrates the density and temparature dependence
of $\zeta$.  (Here we assume that the frequency corresponds to the
$m=2$ $r$-mode frequency of a maximally rotating neutron star:
$\hat\omega = \frac{2}{3}\Omega_{\rm max}$.)  These figures illustrate
the complicated temperature dependence of the visocity due to
superfluid effects.  For temperatures slightly below the superfluid
critical temperature the values of the bulk viscosity are increased
over their normal values.  This is due to an increase in the timescale
$\tau$ which moves it closer to being in resonance with the pulsation
period of the mode.  Once the temperature falls well below the
superfluid critical temparature however, we see that the timescale
$\tau$ becomes even longer than the pulsation period and so the
viscosity becomes smaller again in this case.  We note that even for
very low temperatures there exists a small range of densities, just
above the hyperon threshold densities, where the bulk viscosity
remains rather large.  This is due to the momentum dependence of the
superfluid gap, Eq.~(\ref{lambda_gap}).  The gap $\Delta$ goes to zero
as the Fermi momentum of the particle goes to zero.  Thus just above
the threshold density the superfluid gap vanishes (for any finite
temperatue) so the material in this region will retain the
normal-fluid value of the bulk viscosity.

Our value of $\zeta$ is generally much larger than that obtained
recently by Jones~\cite{jones2}: at a total density
$\rho=7\times10^{14}$g/cm$^3$ and temperature $T=10^{10}$K, our
$\zeta$ is larger than Jones' by a factor of 400. Roughly a factor of
8 is due to the relatively weak coupling we calculate for reaction in
Eq.~(\ref{reacL}). (At this density the $\Sigma^-$ hyperons account
for only about $10\%$ of the bulk viscosity and can be neglected.)  We
note that using the asymptotic values of the weak axial-vector
couplings would erase much of this factor of 8 difference, and thus it
is indicative of the size of the uncertainties in $\zeta$ due to our
poor understanding of nuclear-matter physics. The remaining factor of
50 is thermodynamic in origin. Jones evaluates various partial
derivatives of pressure and chemical potentials, e.g. appearing in
Eq.~(\ref{delta1}) of our paper and Eq.~(42) of Ref.~\cite{jones2},
using the values for a gas of noninteracting fermions.  We include all
the mesonic interaction terms, whose effect is to increase
significantly these thermodynamic derivatives.  Since the details of
the neutron-star equation of state are uncertain, the precise values
of the derivatives are correspondingly uncertain.  However, we think
it unlikely that the true physical mesonic terms will cancel precisely
enough to bring the derivatives down to their noninteracting values.
In summary, we think that the true value of $\zeta$ is within an order
of magnitude of the value we compute here.

\begin{figure}
\includegraphics[width=3.0in]{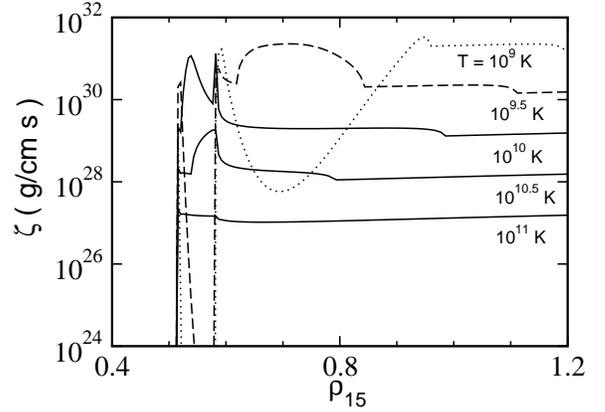}
\caption{Density dependence (in units of $10^{15}$ g/cm${}^3$) of
the hyperon bulk viscosity (in units of g/cm s) for a range
of temperatures.  These curves were constructed using the assumption
$\Delta_{\Sigma^-}=10\Delta_\Lambda$ for the $\Sigma^-$ superfluid gap.
\label{f:zeta10}}
\end{figure}

\section{$r$-mode damping times}
\label{s:rmodes}

The basic formalism for evaluating the effects of bulk viscosity on 
the stability of the $r$-modes is well known~\cite{lom98,lmo99}.  The
time derivative of the co-rotating frame energy $\tilde E$ due to
the effects of bulk viscosity is
\begin{equation}
{d\tilde E\over dt}=-\int {\rm Re}\,\,\zeta\,\,\,
 |\vec\nabla\cdot\delta\vec v|^2
d^{\,3}x.\label{dedt}
\end{equation}
This rotating frame energy $\tilde E$ is (to lowest order in the
angular velocity of the star) given by the integral
\begin{equation}
\tilde E = \frac{1}{2}\int \rho |\delta \vec v|^2 d^{\,3}x.\label{tildeE}
\end{equation}
The timescale $\tau_{B(h)}$ on which hyperon bulk viscosity damps the
mode then is
\begin{equation}
{1\over \tau_{B(h)}}=-{1\over 2\tilde E}{d\tilde E\over dt}. \label{tauB}
\end{equation}
Here we have normalized $\tau_{B(h)}$ so that $1/\tau_{B(h)}$ is the
hyperon bulk viscosity contribution to the imaginary part of the
frequency of the mode.

For the case of the $r$-modes (in slowly rotating stars) the integrals
that determine $\tilde E$ and $d\tilde E/dt$ in Eqs.~(\ref{dedt}) and
(\ref{tildeE}) can be reduced to simple one-dimensional integrals.
For the case of $\tilde E$ this reduction is well known~\cite{lom98}:
\begin{equation}
\tilde E =\frac{1}{2}\alpha^2\Omega^2 R^{-2}\int_0^R \rho r^6 dr.
\label{tildeE_rint}
\end{equation}
Here $\alpha$ represents the dimensionless amplitude of the $r$-mode,
and $\Omega$ and $R$ are the angular velocity and radius of the star
respectively.  The reduction of $d\tilde E/dt$ to a one-dimensional
integral is not so straightforward.  In general the expansion of the
mode $\vec\nabla\cdot\delta\vec v$ is a complicated function of radius
and angle.  To lowest order in slowly rotating stars the bulk
viscosity $\zeta$ will depend only on radius.  Thus we may always
convert Eq.~(\ref{dedt}) to a one-dimensional integral by defining the
angle averaged expansion squared $\langle|\vec\nabla\cdot~\delta\vec
v|^2\rangle$:
\begin{equation}
{d\tilde E\over dt}=-4\pi \int_0^R {\rm Re}\,\,\zeta
\,\,\langle|\vec\nabla\cdot\delta\vec v|^2\rangle r^2dr.\label{dedt_rint}
\end{equation}
While the angle-averaged expansion is in general a complicated
function, for the case of the $r$-modes it is rather simple.  This
function has only been determined numerically~\cite{lmo99}, however
the simple analytical expression,
\begin{equation}
\langle|\vec\nabla\cdot\delta\vec v|^2\rangle=
{\alpha^2\Omega^2\over 690} \left({r\over R}\right)^6
\left[1+0.86\left({r\over R}\right)^2\right],
\label{expansion}
\end{equation}
is an excellent fit to those numerical solutions.

Once the structure of the density function $\rho(r)$ in a stellar
model is determined, it is straightforward to evaluate the integrals
in Eqs.~(\ref{tildeE_rint}) and (\ref{dedt_rint}) using
Eq.~(\ref{expansion}) to determine the bulk viscosity damping time
$\tau_{B(h)}$.  The bulk viscosity of interest to us here is very sensitive
to the density of hyperons in the stellar core.  Thus we use the
relativistic stellar structure equations to evaluate $\rho(r)$.  This
ensures that the size and stucture of the hyperon containing core are
sufficiently accurate for our purposes.  These functions $\rho(r)$ are
illustrated for a range of stellar masses in Fig.~\ref{f:density}
based on the equation of state discussed in Sec.~\ref{s:eos}.  Given
these (numerical) expressions for $\rho(r)$ it is straightforward then
to use the expressions for the hyperon bulk viscosity $\zeta$ derived
in Sec.~\ref{s:times} to obtain $\zeta(r)$ for any given neutron star
temperature.  Together $\rho(r)$ and $\zeta(r)$ then determine
$\tau_{B(h)}$ through Eqs.~(\ref{tildeE_rint}), (\ref{dedt_rint}) and
(\ref{tauB}).  While it is straightforward to evaluate these
timescales, the result is a rather complicated function of the
temperature, angular velocity, and mass of the stellar model and so
we do not attempt to illustrate it directly.

The most important application of the hyperon bulk viscosity timescale
$\tau_{B(h)}$ is the analysis of the role this type of dissipation plays in
the gravitational radiation driven instability in the $r$-modes.
Gravitational radiation contributes a term to the evolution of the
energy $d\tilde E/dt$ that is positive.  As is well known by now,
gravitational radiation tends to drive the $r$-modes unstable in all
rotating stars~\cite{andersson,friedman-morsink}.  As has been
discussed in detail elsewhere~\cite{lom98,lmo99} the imaginary
part of the frequency of the $r$-mode may be written as:
\begin{equation}
{1\over\tau_r}=-{1\over\tau_{GR}}+{1\over\tau_{B(h)}}+{1\over\tau_{B(u)}}.
\label{totaltau}
\end{equation}
Here $\tau_{GR}$ represents the timescale for gravitational radiation
to effect the $r$-mode, $\tau_{B(h)}$ is the hyperon bulk viscosity
timescale discussed here, and $\tau_{B(u)}$ is the modified Urca bulk
viscosity.  Detailed expressions for evaluating these other terms are
discussed elsewhere and will not be repeated here.  Suffice it to say
that each is a function of the temperature, angular velocity and mass
of the neutron star.  Since $1/\tau_r$ is the imaginary part of the
frequency of the $r$-mode, the mode is stable when $\tau_r>0$ and
unstable when $\tau_r<0$.  For a star of given temperature and mass,
the critical angular velocity $\Omega_c$ is defined to be the angular
velocity where $1/\tau_r=0$.  Stars rotating more rapidly than
$\Omega_c$ are unstable while those rotating more slowly are stable.

\begin{figure}
\includegraphics[width=3.0in]{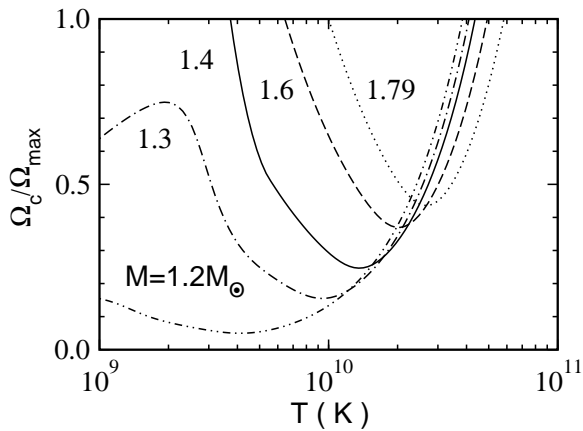}
\caption{Critical angular velocities for neutron stars as a function of
hyperon core temperature.  Each curve represents a neutron star of fixed
mass, ranging from 1.2 $M_\odot$ to the maximum mass for this equation of
state, 1.79 $M_\odot$.  These curves assume $\Delta_\Sigma=\Delta_\Lambda$
and use the $\beta$-decay values of the weak axial vector coupling
coefficients.\label{f:om_crit_m}}
\end{figure}

\begin{figure}
\includegraphics[width=3.0in]{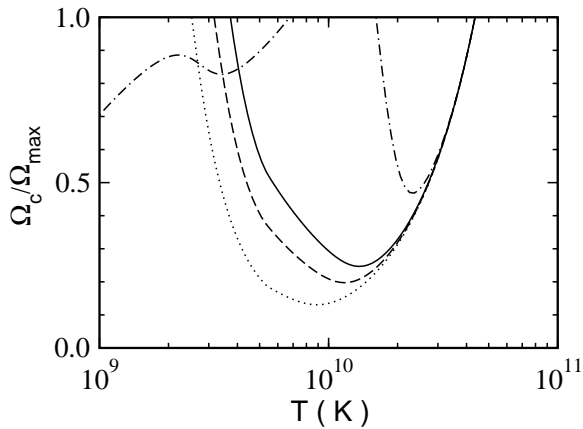}
\caption{Critical angular velocities for 1.4 $M_\odot$ neutron stars
as a function of hyperon core temperature.  The solid curve assumes
that $\Delta_\Sigma = \Delta_\Lambda$ while the dot-dash curve assumes
$\Delta_\Sigma = 10\Delta_\Lambda$.  Both curves use the $\beta$-decay
values of the weak axial vector coupling coefficients.  Dotted curve
uses $g_{B}=-1$, while dashed curve uses bare masses.
\label{f:om_crit_m1.4_gall}}
\end{figure}

We have evaluated the critical angular velocities $\Omega_c$
numerically using the new hyperon bulk viscosities derived in
Sec.~\ref{s:times}.  Figure~\ref{f:om_crit_m} illustrates the
temperature dependence of the critical angular velocities for a range
of neutron-star masses.  The more massive neutron stars have larger
hyperon cores which suppress the $r$-mode instability more
effectively.  The curves in Fig.~\ref{f:om_crit_m} assume that the
$\Sigma^-$ superfluid gap function is given by
$\Delta_\Sigma=\Delta_\Lambda$, and that the axial vector coupling
coefficients have their $\beta$-decay values.  In
Fig.~\ref{f:om_crit_m1.4_gall} we compare the critical angular
velocity curves for 1.4$M_\odot$ stellar models using either the
$\Delta_\Sigma=\Delta_\Lambda$ (solid curve) or the
$\Delta_\Sigma=10\Delta_\Lambda$ (dash-dot curve) assumption about the
$\Sigma^-$ superfluid gap.  The larger value of $\Delta_\Sigma$ allows
superfluidity to make the bulk viscosity larger over a wider range of
temperatures, and hence the $r$-mode instability is less effective.
Also illustrated in Fig.~\ref{f:om_crit_m1.4_gall} are the effects of
changing various microphysics assumptions.  The dotted curve shows the
effect of changing the values of the axial vector coupling constants
from their $\beta$-decay values to the asymptotic value -1.  And the
dashed curve shows the effect of using bare masses rather than the
effective masses in the scattering cross sections.  These alternative
assumptions make the bulk viscosity less effective and the $r$-mode
instability operates over a wider range of angular velocities in these
stars. However, neither of these effects is as large as that resulting
from a change in the superfluid gap.

\begin{figure}
\includegraphics[width=3.0in]{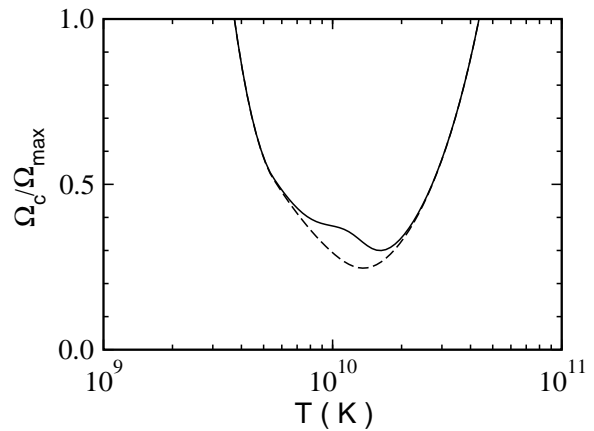}
\caption{Critical angular velocities for 1.4$M_\odot$ neutron stars as
a function of hyperon core temperature.  Solid curve includes the
effects of hyperon bulk viscosity, hyperon channel direct Urca bulk
viscosity, and modified Urca bulk viscosity.  Dashed curve leaves out
the effects of the direct Urca hyperon bulk
viscosity.\label{f:om_crit_m1.4_gn_direct}}
\end{figure}

The hyperons' primary contribution to the bulk viscosity of neutron-star
matter is through the mechanism discussed above.  However, as pointed out
by Jones, the presence of hyperons in the core of a neutron star also make
it possible for alternate forms of the direct Urca interaction to
take place and these too contribute to the bulk viscosity of the material.
Jones showed that the contributions to the bulk viscosity from this
process are given by
\begin{eqnarray}
{\rm Re}\,\,\zeta&=&{4.9\times 10^{30} T_{10}^{-4}\over 1+
2.0\times 10^{-6}\hat\omega^2T_{10}^{-8}},
\label{e:direct_urca}
\end{eqnarray}
in cgs units for typical values of neutron star matter, where $T_{10}$
is the temperature measured in units of $10^{10}$K.  Using this
expression in the $\Lambda$ containing core of the neutron
star, we have evaluated the effects of this hyperon channel direct
Urca bulk viscosity on the stability of the $r$-modes.  These results
are illustrated in Fig.~\ref{f:om_crit_m1.4_gn_direct}.  The solid
curve in Fig.~\ref{f:om_crit_m1.4_gn_direct} includes the effects of
the hyperon bulk viscosity discussed above, the hyperon channel
direct Urca bulk viscosity of Eq.~(\ref{e:direct_urca}), and the ordinary
modified Urca bulk viscosity.  For comparison the dashed curve leaves
out the effects of the hyperon channel direct Urca bulk viscosity.
We see that this direct Urca bulk viscosity has only a small effect on
the stability of the $r$-modes for temperatures around $10^{10}$K.

\section{Discussion}
\label{s:end}

We have analyzed here the effects of the bulk viscosity due to
hyperons on the stability of the $r$-modes in rotating neutron stars.
Hyperons exist only in the high density core of a neutron star where
the influence of the $r$-mode is quite small.  Thus to evaluate
accurately and reliably the importance of this effect, it was
necessary to compute detailed and accurate models of the composition
and structure of the neutron star core, and to have an accurate model
of the structure of the $r$-mode in this region.  We use
Glendenning's~\cite{glendpaper,glendbook} relativistic mean-field
equation of state to evaluate the composition of the nuclear matter in
the stellar core, and solve the relativistic Oppenheimer-Volkoff
equations to determine the stellar structure. 

Our evaluation of the
hyperon bulk viscosity improves on previous work in several ways.
First we generalize in Eqs.~(\ref{zeta}) and (\ref{zeta2}) the
standard expression for the bulk viscosity coefficient so that it
applies to relativistic fluids such as neutron star matter.  Second we
generalize the expressions in Eqs.~(\ref{delta1}) and (\ref{delta2})
for the thermodynamic quantities that relate the microscopic reaction
rates to the relaxation time that appears in the expression for the
bulk viscosity by including fully interacting nuclear matter.  And
third we obtain in Eqs.~(\ref{m2lambda}) and (\ref{m2sigma}) the fully
relativistic expressions for the relevant hyperon scattering cross
sections needed to evaluate the microscopic reaction rates.  While our
expressions for these cross sections reduce to the published
low-momentum results, we find that the difference of Eq.~(\ref{m2sigma})
from the low-momentum limit can be an order of magnitude or more and reduces
the coefficient of bulk viscosity somewhat at low densities.  Finally we
evaluate the effects of this hyperon bulk viscosity on the $r$-modes using
an accurate model for the structure of the $r$-mode in the core of a neutron
star.

Our results show that the hyperon bulk viscosity does not substantially
suppress the gravitational radiation instability of the $r$-modes until the
temperature of the core of the neutron star drops below a few times $10^9$K.
(This is in spite of the fact that our coefficient of bulk viscosity is
actually higher than that of Jones~\cite{jones01,jones2}. The expansion of
the fluid in the core of the star as given in Eq.~\ref{expansion} is
smaller than he estimated.) Below $10^9$K the $r$-mode instability is
strongly suppressed in all of our models over the range of $\Sigma^-$
superfluid gap functions and the range of axial vector coupling constants
that we studied.  If the core of the neutron star cools according to the
standard modified Urca process~\cite{shapiro-teuk}, then it would remain hot
enough for the $r$-mode instability to act for about a day.  This is enough
time for the $r$-mode to grow and radiate away through gravitational waves a
substantial fraction of the rotational kinetic energy of a rapidly rotating
neutron star~\cite{ltv}.  However if the core of the neutron star cools
substantially faster than this, then it may not be possible for the $r$-mode
to grow rapidly enough to effect the star in a substantial way before the
hyperon bulk viscosity stabilizes it.  Cooling by the direct Urca process is
significantly faster that the modified Urca process: cooling the core of a
neutron star to a few times $10^9$K within about a
second~\cite{pt,page-etal}.  Cooling by the direct Urca process will occur
in neutron-star matter whenever the proton/baryon ratio is larger than about
0.15.  Since proton fractions in excess of this are now generally expected
in neutron star matter, cooling by the direct Urca process seems likely at
least until the temperature of the core falls below the superfluid
transition temperature for neutrons or protons at about $10^9$K.  Thus it
appears likely that the $r$-mode instability is effectively suppressed by
rapid cooling of the neutron star core and the non-leptonic hyperon bulk
viscosity.

Once a neutron star cools below the transition temperature for the formation
of neutron and proton superfluids, the relaxation timescale for the hyperon
interactions will increase exponentially compared to the expressions derived
here in Eqs. (\ref{taus}) and (\ref{tausl}).  This sharply reduces via
Eq.~(\ref{realzeta}) the bulk viscosity from this process at sufficiently
low temperatures.  Further detailed calculations would be needed to
determine whether the hyperon bulk viscosity has a significant influence on
the $r$-mode instability at temperatures of a few times $10^8\,$K, which are
expected to exist in the cores of neutron stars in low-mass x-ray binaries.
We did not carry out these calculations in part because solid crust-related
shear~\cite{bu,lou} and magnetic field~\cite{mendell} effects are quite
effective in suppressing the instability at these low temperatures.

How robust is the conclusion that the $r$-mode instability is effectively
suppressed?  Clearly the details of the nuclear physics
involving hyperons in neutron star matter are not well understood at
this time.  However, our conclusion applies to the entire expected
range of the most poorly known properties of this material: the
superfluid $\Sigma^-$ gap function and the axial vector coupling
coefficients.  In order to escape this conclusion, it would be
necessary for neutron star matter to have very few hyperons present at
the densities which exist in the cores of real neutron stars.  This
would require the equation of state to be substantially different from
the one studied here, or the masses of neutron stars to be
significantly smaller than $1.4M_\odot$.  Rapid rotation also lowers
the central density and consequently the size of the hyperon core in a
neutron star.  The central density of a maximally rotating
$1.4M_\odot$ neutron star is about 73\% of its non-rotating value (for
the equation of state studied here)~\cite{rns}.  This reduction almost
eliminates the hyperon core for this extreme angular velocity, but
over almost all of the range of angular velocities, 1.4$M_\odot$ stars
have substantial hyperon cores.  Finally if the dissipation in the
core were sufficiently large it might be possible for the $r$-mode
eigenfunction to be clamped to zero in the core by the dissipative
processes while remaining finite and unstable in the outer parts of
the star.  The discussion of this possibility in the appendix shows
that the hyperon bulk viscosity is not strong enough to clamp the
$r$-mode in this way.

\begin{acknowledgments}
We thank P. B. Jones for raising the issue of hyperon bulk viscosity,
for communicating his unpublished work to us, and for giving us
helpful comments on our work.  We also thank S.  Balberg, D. Chernoff,
J. Friedman, N. Glendinning,
P. Goldreich, J. Lattimer, G. Mendell, J. Miller, E. S.
Phinney, M. Prakash, R. Sawyer, and K. Thorne for helpful discussions. This
work was supported by NSF grants PHY-9796079, PHY-0071028, PHY-0079683, and
PHY-0099568, and NASA grants NAG5-4093 and NAG5-10707.
\end{acknowledgments}

\begin{appendix}
\section{Mode clamping}

Bulk viscosity damps a mode by dissipating energy according to the
expression
\begin{equation}
{d\tilde E\over dt}= - {\tilde E\over 2 \tau_B}.
\end{equation}
For the case of the $r$-modes in slowly rotating stars, we may
express the energy, and its time derivative as simple radial integrals:
\begin{equation}
\tilde E =\int \epsilon dr,
\end{equation}
\begin{equation}
{d \tilde E\over dt} = -\int \dot\epsilon dr,
\end{equation}
where $\epsilon$ and $\dot\epsilon$ are the angle averaged
energy and energy dissipation rate densities respectively as given in
Eqs.~(\ref{tildeE_rint}) and (\ref{dedt_rint}). 

The mode will be completely suppressed (clamped) locally if the amount
of energy removed from the mode locally in one oscillation period is
comparable to the local energy density of the mode.  Thus we define
the local quality factor of the mode:
\begin{equation}
q = {\hat\omega \epsilon\over 2\pi \dot\epsilon}.
\end{equation}
If $q\lesssim 1$ the mode will be clamped.  For the $r$-modes
we find that
\begin{equation}
{1\over q} \approx  {0.3\zeta\over \rho R^2 \Omega_{\rm max}}
\left({r\over R}\right)^2{\Omega_{\rm max}\over\Omega}.
\end{equation}
For the $1.4M_\odot$ neutron star model considered here
$\Omega_{\rm max}\approx 4700$ rad/s, $\rho\gtrsim 5\times
10^{14}$ g/cm${}^3$ and $r\lesssim 6$ km in the region where hyperons
occur, and $R\approx 14$ km.  Thus
\begin{equation}
{1\over q} \approx {\zeta\over 8\times 10^{31}} {\Omega_{\rm max}\over
\Omega}
\end{equation}
for this case.  Figures~\ref{f:zeta} and \ref{f:zeta10} show that the
bulk viscosity never exceeds about $10^{31}$ in the temperature range
of interest to us.  Thus, we conclude that the hyperon bulk viscosity
represents a small perturbation on the basic hydrodynamic forces of
the $r$-modes.  The condition $q<1$ is violated only for relatively
slowly rotating stars.  In the domain where the gravitational
radiation instability is most likely to be important, the dissipation
by hyperons represents a small perturbation on the basic hydrodynamic
forces, thus the $r$-modes will not be clamped.
\end{appendix}

\end{document}